\documentclass[cernpreprint, atlasdraft=false, texlive=2023, UKenglish, texmf, orcidlogo]{atlasdoc}
 
\usepackage{atlaspackage}
\usepackage{atlasbiblatex}
 
\usepackage{atlasphysics}

\graphicspath{{logos/}{figures/}}
 
\usepackage{ANA-HDBS-2019-09-PAPER-defs}

\AtlasTitle{Search for the decay of the Higgs boson to a $Z$ boson and a
light pseudoscalar particle decaying to two photons}

\AtlasAbstract{
A search for the decay of the Higgs boson to a $Z$ boson and a light, pseudoscalar particle, $a$, decaying respectively to two leptons and to two photons is reported.
The search uses the full LHC Run 2 proton--proton collision data
at $\sqrt{s}=13$ TeV,
corresponding to  139~fb$^{-1}$  collected by the ATLAS detector.
This is one of the first searches for this specific decay mode of the
Higgs boson, and it probes
unexplored parameter space in models with axion-like particles (ALPs)
and extended scalar sectors.
The mass of the $a$ particle is assumed to be in the range 0.1--33 GeV. The data are analysed in two categories: a merged category where the photons from the $a$ decay are reconstructed in the ATLAS calorimeter as a single cluster, and a resolved category in which two separate photons are detected. The main background processes are from Standard Model $Z$ boson production in association with photons or jets.
The data are in agreement with the background predictions, and upper limits on
the branching ratio of the Higgs boson decay to $Za$ times the branching ratio $a\to\gamma\gamma$
are derived at the 95\% confidence level and they range from 0.08\% to 2\% depending on the mass of the $a$ particle. The results are also interpreted in the context of ALP models.
}

\AtlasRefCode{HDBS-2019-09}
 
\PreprintIdNumber{CERN-EP-2023-167}

\AtlasJournalRef{\PLB 848 (2024) 138536}
\AtlasDOI{10.1016/j.physletb.2024.138536}

\hypersetup{pdftitle={ATLAS document},pdfauthor={The ATLAS Collaboration}}
 
\begin{document}
 
\maketitle

\section{Introduction}
\label{sec:intro}

Light pseudoscalars that couple to Higgs bosons appear in many
well-motivated new physics scenarios. A few examples of such scenarios are
the next-to-minimal supersymmetric standard model~\cite{Maniatis:2009re},
axion models~\cite{Bauer:2017nlg,Bauer:2017ris},
and models with dark or extended Higgs sectors~\cite{Curtin:2013fra,Robens:2015gla,Yang:2018fje}.
In addition, they may be related to a possible muon $g-2$ deviation with respect to the
Standard Model (SM) predicted value, e.g. as in Ref.~\cite{PhysRevD.94.115033}.
The search for elusive particles in the decays of the 125~\GeV\ Higgs boson~\cite{HIGG-2012-27,CMS-HIG-12-028}
is a highly promising avenue for their discovery. Due to the narrow decay
width of the Higgs boson~\cite{LHCHiggsCrossSectionWorkingGroup:2011wcg},
even minor contributions from physics beyond the SM can result in final
states with significant branching fractions.
The most recent combination of the ATLAS measurements of the Higgs
boson production rates does not rule out
decays to yet unobserved states with a branching ratio
of up to 12\%~\cite{HIGG-2021-23} at
95\% confidence level (CL). The corresponding result from
the CMS experiment is 16\%~\cite{CMS-HIG-22-001}. These results
indicate that there may be sizeable exotic decays of the Higgs bosons
that have not been observed yet.
The LHC experiments have looked for such particles in Higgs boson
decays, usually assuming that the Higgs boson, $H$, decays to a pair
of narrow-width, pseudoscalar particles $a$ ($H\to aa$), which
subsequently decay to fermions or photons.
Such searches have been performed in the
$bb\mu\mu$ \cite{HDBS-2021-03,CMS-HIG-18-011},
$4b$~\cite{HIGG-2017-05, HDBS-2018-47},
$4\mu$~\cite{CMS-HIG-18-003,EXOT-2016-22},
$\mu\mu\tau\tau$ or $4\tau$~\cite{CMS-HIG-18-024,CMS-HIG-18-006,CMS-HIG-17-029,CMS-HIG-16-015,HIGG-2014-02,CMS-HIG-17-024},
$bb\tau\tau$~\cite{CMS-HIG-20-014},
$4\gamma$~\cite{EXOT-2013-24,CMS-HIG-21-003,CMS-HIG-21-016}
and $2\gamma$ + 2 jets \cite{HIGG-2017-09} final states.
In addition to the $H\to aa$ channel, the decay mode of $H\to Za$
is well-motivated, but remains less studied.
Such decays are particularly motivated by
axion models~\cite{Bauer:2017nlg}. This decay has been searched for
by ATLAS in the case where the $Z$ boson decays to leptons
and $a$ decays to jets~\cite{HDBS-2018-37}, but this is one of the first
searches for the $H\to Za$ with $a\to\gamma\gamma$.
In this letter, results
from the search for the final state in which the $Z$ boson decays to leptons
and the $a$ particle decays to a photon pair are presented using the full
LHC Run 2 proton--proton ($pp$) collision data collected by the ATLAS
detector during 2015--2018 and corresponding to an integrated luminosity of 139~fb$^{-1}$.
The search explores $a$ particle masses, $m_a$, between 0.1~\GeV\ and \mamax~\GeV.
When the invariant mass of the two photons from the $a$ decay is below 2~\GeV\
and the transverse momentum of the diphoton system is high,
their electromagnetic
showers typically overlap in the detector and a single photon is reconstructed.
The search presented here considers both the merged and the resolved
$a\to \gaga$ decays.
 
This letter is structured as follows.
A description of the ATLAS detector is given in
Section~\ref{sec:detector}, followed by a presentation of the
data and simulated samples in Section~\ref{sec:samples}. The
physics object reconstruction is discussed in Section~\ref{sec:objects},
followed by an explanation of the analysis in Section~\ref{sec:analysis},
the presentation  of the search results in Section~\ref{sec:results}
and the conclusion in Section~\ref{sec:conclusion}.


\section{ATLAS detector}
\label{sec:detector}

The ATLAS detector~\cite{PERF-2007-01}
at the LHC is a multipurpose particle detector with a forward-backward
symmetric cylindrical geometry and a near 4$\pi$ coverage in solid angle.\footnote{ATLAS
uses a right-handed coordinate system with its origin at the
nominal interaction point (IP) in the centre of the detector
and the $z$-axis along the beam pipe. The $x$-axis points from the IP
to the centre of the LHC ring, and the $y$-axis points upwards. Cylindrical
coordinates $(r, \phi)$ are used in the transverse plane,
$\phi$ being the azimuthal angle around the $z$-axis.
The pseudorapidity is defined in terms of the polar angle
$\theta$ as $\eta = - \ln\tan(\theta/2)$.
Angular distance is measured in units of
$\Delta R = \sqrt{ (\Delta\eta)^2 + (\Delta\phi)^2 }$.}
It consists of an inner tracking detector surrounded by a thin
superconducting solenoid providing a 2~T axial magnetic field,
electromagnetic (EM) and hadronic calorimeters, and a muon spectrometer.
The inner tracking detector covers the pseudorapidity
range $|\eta| < 2.5$. It consists of silicon pixel, silicon microstrip,
and transition radiation tracking detectors.
Lead/liquid-argon (LAr) sampling calorimeters provide EM
energy measurements with high granularity. A steel/scintillator-tile
hadronic calorimeter covers the central pseudorapidity range
($|\eta| < 1.7$). The endcap and forward regions are instrumented
with LAr calorimeters for both the EM and hadronic energy measurements
up to $|\eta| = 4.9$.
The muon spectrometer surrounds the calorimeters
and is based on three large superconducting air-core toroidal magnets with
eight coils each. The field integral of the toroids ranges between
2.0 and 6.0~T$\cdot$m across most of the detector. The muon spectrometer
includes a system of precision chambers for tracking and fast detectors for triggering. A two-level trigger system is used to select
events~\cite{TRIG-2016-01}.
The first-level trigger is implemented in hardware and uses a subset of
the detector information to accept events at a rate below 100 kHz.
This is followed by a software-based
trigger that reduces the accepted event rate to 1~kHz on average.
An extensive software suite~\cite{ATL-SOFT-PUB-2021-001}
is used in the reconstruction
and analysis of real and simulated data, in
detector operations, and in the trigger and data acquisition
systems of the experiment.


\section{Data and simulated event samples}
\label{sec:samples}
 
The search for $H\to Za$ is performed using the full LHC Run 2 (2015--2018)
$\sqrt{s}= 13$~\TeV\ $pp$ collision dataset recorded
by the ATLAS detector and corresponding to an integrated
luminosity of $139.0$~fb$^{-1}$~\cite{DAPR-2021-01},
which includes only data-taking periods where all relevant detector
subsystems were operational~\cite{DAPR-2018-01}.
The data events are collected using a set of
triggers that require the presence of one or two
electrons~\cite{TRIG-2018-05}  or one or two  muons~\cite{TRIG-2018-01}
with thresholds in the transverse energy
for electrons and the transverse momentum for muons that are
in the range 12--26~\GeV, depending on the lepton flavour, the number of leptons
and the data-taking period~\cite{TRIG-2019-03}.
Data-driven corrections are applied to the event-level trigger efficiencies.
 
Simulated event samples are used to optimise the analysis selections and
characterise the signal and backgrounds. The signal samples consist of events
with Higgs bosons produced via gluon--gluon fusion or vector-boson-fusion
(VBF). Additional Higgs boson production mechanisms are found to have
a negligible impact and thus not considered.
Higgs boson production via gluon--gluon fusion is simulated at next-to-next-to-leading-order (NNLO) accuracy in
QCD using \POWHEGBOX[v2]~\cite{Hamilton:2013fea,Hamilton:2015nsa,Alioli:2010xd,Nason:2004rx,Frixione:2007vw}.
The simulation achieves NNLO accuracy for arbitrary inclusive \(gg\to H\) observables by reweighting the Higgs boson
rapidity spectrum in \textsc{Hj}-\MINLO~\cite{Hamilton:2012np,Campbell:2012am,Hamilton:2012rf} to that of HNNLO~\cite{Catani:2007vq}.
Higgs boson production via VBF is simulated at next-to-leading order (NLO) precision using \POWHEGBOX[v2].
The \PDFforLHC[15nnlo] PDF set~\cite{Butterworth:2015oua} is used for both production mechanisms.
The parton showering and hadronisation is performed using \PYTHIA[8.244.3]~\cite{Sjostrand:2014zea}
with parameters set according to the \AZNLO set of tune parameters (tune)~\cite{STDM-2012-23}.
The simulated Higgs production samples are normalised using
dedicated cross-section calculations. For gluon--gluon fusion
the next-to-next-to-next-to-leading-order cross-section in QCD plus
electroweak corrections at
NLO~\cite{deFlorian:2016spz,Anastasiou:2016cez,Anastasiou:2015ema,Dulat:2018rbf,Harlander:2009mq,Harlander:2009bw,Harlander:2009my,Pak:2009dg,Actis:2008ug,Actis:2008ts,Bonetti:2018ukf,Bonetti:2018ukf} is used.
For VBF an approximate NNLO QCD cross-section with NLO electroweak
corrections~\cite{Ciccolini:2007jr,Ciccolini:2007ec,Bolzoni:2010xr}
is employed.
 
The Higgs bosons that are generated undergo subsequent decay into a $Z$ boson and a pseudoscalar particle, $a$,
through \PYTHIA[8.244.3]~\cite{Sjostrand:2014zea}. The decay of the $Z$ boson is isotropic in its decay frame. The simulated samples are generated with
the $a$ particle mass in the range 0.1--\mamax~\GeV.

The main backgrounds for this search come from $Z$ boson production in
association with jets ($Z$+jets) or a photon ($Z+\gamma$).
The production of $Z$+jets is simulated  with the
\SHERPA[2.2.1]~\cite{Bothmann:2019yzt} generator using NLO matrix elements for up to two partons, and leading-order (LO) matrix elements
for up to four partons calculated with the Comix~\cite{Gleisberg:2008fv}
and \OPENLOOPS~\cite{Buccioni:2019sur,Cascioli:2011va,Denner:2016kdg} libraries. They
are matched with the \SHERPA parton shower~\cite{Schumann:2007mg} using the \MEPSatNLO
prescription~\cite{Hoeche:2011fd,Hoeche:2012yf,Catani:2001cc,Hoeche:2009rj}
using the set of tuned parameters developed by the \SHERPA authors.
The \NNPDF[3.0nnlo] set of PDFs~\cite{Ball:2014uwa} is used and the samples
are normalised to a NNLO prediction~\cite{Anastasiou:2003ds}.
The production of ($Z+\gamma$) final states is simulated with the
\SHERPA[2.2.8] generator.  Matrix elements at
NLO QCD accuracy for up to one additional parton and LO accuracy for
up to three additional parton emissions are matched and merged with
the \SHERPA parton shower based on Catani--Seymour
dipole factorisation~\cite{Gleisberg:2008fv,Schumann:2007mg} using the MEPS@NLO
prescription.
The virtual QCD corrections for matrix elements at NLO accuracy are provided by
the \OPENLOOPS[2] library~\cite{Buccioni:2019sur,Cascioli:2011va,Buccioni:2017yxi,Denner:2016kdg}.
The generation uses the \NNPDF[3.0nnlo] PDF set~\cite{Ball:2014uwa},
along with the dedicated set of tuned parton-shower parameters developed by
the \SHERPA authors.
Finally, the Higgs boson decay $H\to Z\gamma$~\cite{HIGG-2016-14}
is found to be negligible, and thus not included.
 
The effects of multiple $pp$ interactions in the same bunch crossing as the
hard scatter, and in neighbouring bunch crossings, (pile-up) are included using simulated events
generated with \PYTHIA[8.186]~\cite{Sjostrand:2007gs} with the A3 minimum-bias tune. Simulated events
are weighted to
reproduce the distribution of the average number of interactions per
bunch crossing observed in data.
 
All the generated events are passed through the simulation of the
ATLAS detector~\cite{SOFT-2010-01} based on Geant4~\cite{Agostinelli:2002hh} and
the same algorithms as for data.
 
\section{Object reconstruction}
\label{sec:objects}
 
Electron candidates are created from an energy deposit in the EM
calorimeter matched with a track in the inner detector. They are required
to be within the fiducial region of the detector, $ |\eta| < 2.47$, excluding
the transition region between the barrel and endcap calorimeters
($1.37 < |\eta| < 1.52$). The associated track must have
$|d_0 |/\sigma_{ d_ 0} < 5$ and $|z_0 | \sin \theta < 0.5$~mm, where $d_0$ ($z_0$)
is the transverse (longitudinal) impact parameter
relative to the primary vertex, and $\sigma_{d_0}$ is the uncertainty in $d_0$.
The reconstructed electrons are
required to have $\pt > 20$~\GeV\ and their quality is ensured by using
a likelihood method with the requirement to satisfy ``medium''
identification criteria~\cite{EGAM-2018-01}.
The energy of the electron is calibrated through a series of
successive steps, utilising a combination of
simulation-based and data-driven correction factors~\cite{EGAM-2018-01}.
 
Muon candidates are reconstructed in the range $|\eta| < 2.5$ by combining
tracks in the inner detector with tracks in the
muon spectrometer. Muon candidates must have
$|d_0 |/\sigma_{ d_ 0} < 3$ and $|z_0 | \sin \theta < 0.5$~mm.
In addition they must have $\pt > 20$~\GeV~and satisfy
``medium'' quality requirements~\cite{PERF-2015-10}.
The efficiency of muon reconstruction and identification,
along with the momentum calibration and associated systematic uncertainties,
have been estimated using the methods described in Refs~\cite{PERF-2015-10,MUON-2018-03}.
 
Both muon and electron candidates are required to satisfy isolation requirements
based on tracking and calorimeter measurements in order to suppress hadronic
and non-prompt lepton backgrounds~\cite{EGAM-2018-01,MUON-2018-03}.
In addition, electrons reconstructed within $\Delta R = 0.2$ from a muon
are not considered in the analysis.
 
Data-driven corrections are applied at the event level to ensure accurate reconstruction,
identification and isolation efficiencies of muons and electrons.
 
Photon candidates are reconstructed from topological clusters of energy
deposits in the EM calorimeter and calibrated as described in
Ref.~\cite{EGAM-2018-01}. Photon candidates are required to
have $E_\mathrm{T} > 10$~\GeV\ and $|\eta| < 2.37$, excluding the transition region between
the barrel and endcap calorimeters.
Backgrounds from jets are reduced by requiring them to fulfil  ``loose''  identification
criteria based on shower shapes in the ATLAS EM calorimeter~\cite{EGAM-2018-01}.
Finally, photon candidates are not considered if they are
reconstructed within $\Delta R < 0.3$ from an electron or muon.


\section{Analysis}
\label{sec:analysis}

The signature that is searched for consists of two high-\pt\
leptons from the $Z$ boson decay and a  photon pair from the
$a$ particle decay.
The boost of the photon pair determines whether it will be reconstructed as a
single photon or as two resolved photons.
This motivates a categorisation of events depending on whether the $\gaga$
pair is reconstructed as one (``merged'' category)
or two (``resolved'' category) photons.
The main sources of background for both categories is $Z+$jets and \Zga
production.
 
The events used in this search are required to have exactly two electrons
or two muons (hereafter referred to as the two leptons),
with one of them satisfying $\pt > 27$~\GeV\ and the other $\pt > 20$~\GeV.
Subsequently,
the two leptons are required to have opposite electric charge
and their invariant mass must be in the range 81--101~\GeV\ to be
compatible with the mass of the $Z$ boson.
Finally, the transverse momentum of the two-lepton system is
required to be greater than 10~\GeV, a criterion that is optimised
to improve both background rejection and to reduce uncertainties
in the background modelling.
 
The events are further required to contain at least one photon.
If the event contains two or more photons, for all  possible
pairs for which the angular distance between the two photons
satisfies $\Delta R_{\gamma\gamma} < 1.5$,
the quantity $X = \Delta R_{\gamma\gamma} \pt^{\gamma\gamma} /(2m_{\gamma\gamma}) $ is  calculated.
In this variable, $\pt^{\gamma\gamma}$ and $m_{\gamma\gamma}$ denote the transverse momentum
and the invariant mass of the photon pair, respectively.\footnote{For a
highly boosted pair
the angular distance between the objects is approximately given by
$\Delta R = 2m/\pt$, where $m$ and $\pt$ are the mass and the transverse
momentum of the pair. Therefore the variable $X$ can help
to assess the compatibility
of the di-photon candidate with a boosted $a\to\gamma\gamma$ decay
and in turn to select the best such candidate,
if there is more than one photon pair.}
If there is a photon pair with $0.96 <X< 1.2$ then the event is
included in the ``resolved'' category. In case of
multiple photon pairs satisfying this requirement, the pair
with the $X$ value closest to unity is considered as
the $a\to \gamma\gamma$ pair candidate.
If the event fails the resolved
category requirement, then it becomes part of the ``merged'' category
if it contains at least one photon with $\pt > 20$~\GeV. In this case,
the highest-\pt photon is selected as the $a\to \gamma\gamma$ decay candidate.
The resolved category is sensitive to relatively high $a$ particle masses,
whereas the merged category is designed for the low mass regime with typically $m_a \lesssim 2$~\GeV.

After event categorisation, the $a\to \gamma\gamma$ decay candidate
photons are required to pass isolation criteria~\cite{EGAM-2018-01}.
For the merged category, only a track based isolation criterion is used by
requiring the sum of the \pt of the reconstructed tracks within a cone of
$\Delta R = 0.2$ around the photon to be less than 5\% of the photon \pt. For
the resolved category photons both track and calorimetric isolation are used.
For the track isolation, if $\Delta R(\gamma,\gamma) > 0.22$ the same
track-based photon isolation is used as for the merged category;
otherwise no track isolation is applied.
For the calorimeter isolation, the sum of the topological clusters \Et
in a cone size $\Delta R = 0.2$ around the photon is required to
be less than 6.5\% of the photon \pt.
The isolation cuts efficiency exceeds 94\% and 51\% for isolated photons that
pass the identification criteria discussed in Section~\ref{sec:objects} and the merged
and resolved selection respectively.
 
The total event selection acceptance times efficiency for signal events in the full mass range
in the merged and resolved categories is in the range 4--11\%
and 0.4--2\%, respectively for $(H\to Za\to \ell\ell\gamma\gamma)$ generated events.

\subsection{Resolved category}\label{subsec:resolved}
 
\begin{figure}
\centering
\subfloat[]{ \label{fig:resolved:mllgg}
\includegraphics[width=0.45\textwidth]{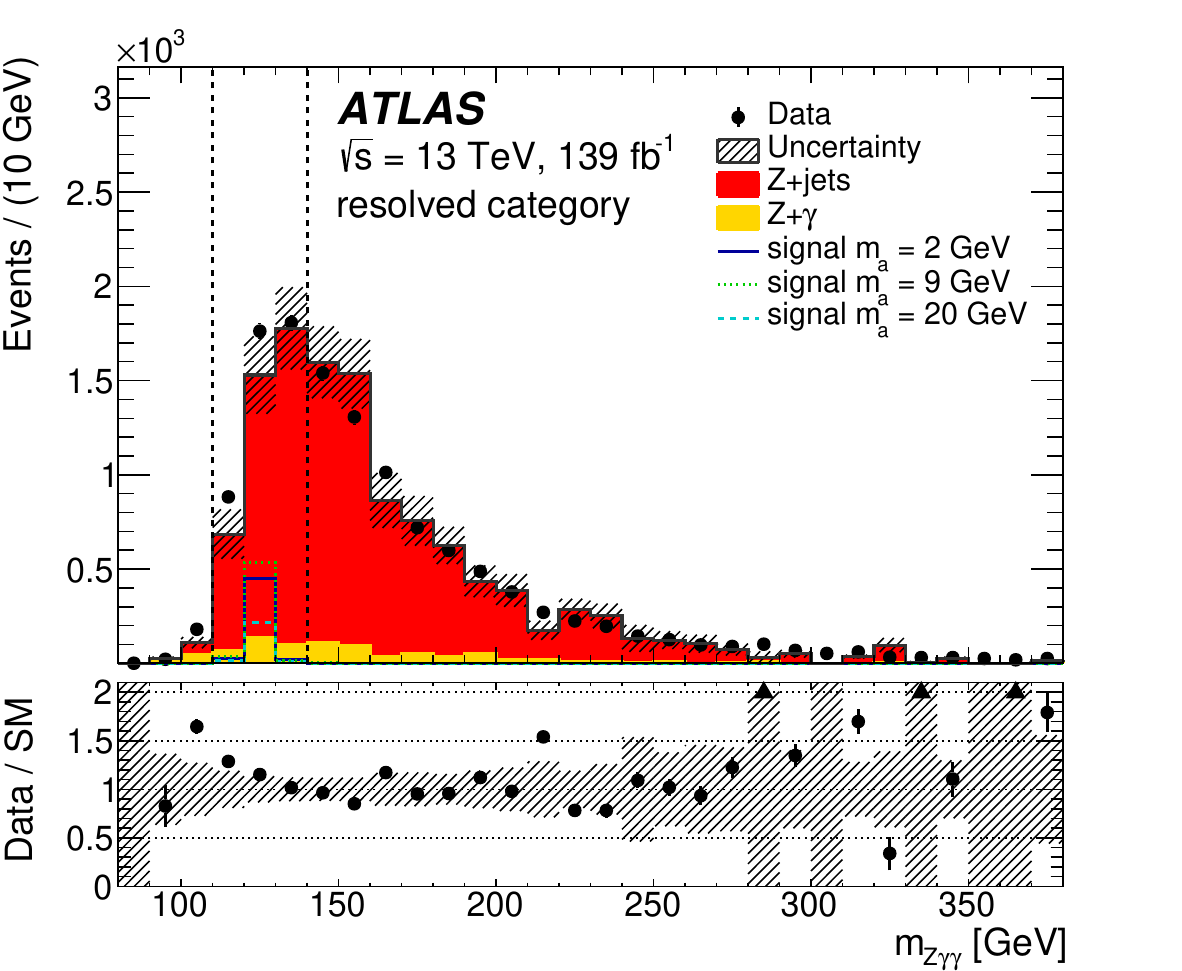}}
\subfloat[]{\label{fig:resolved:signal}
\includegraphics[width=0.45\textwidth]{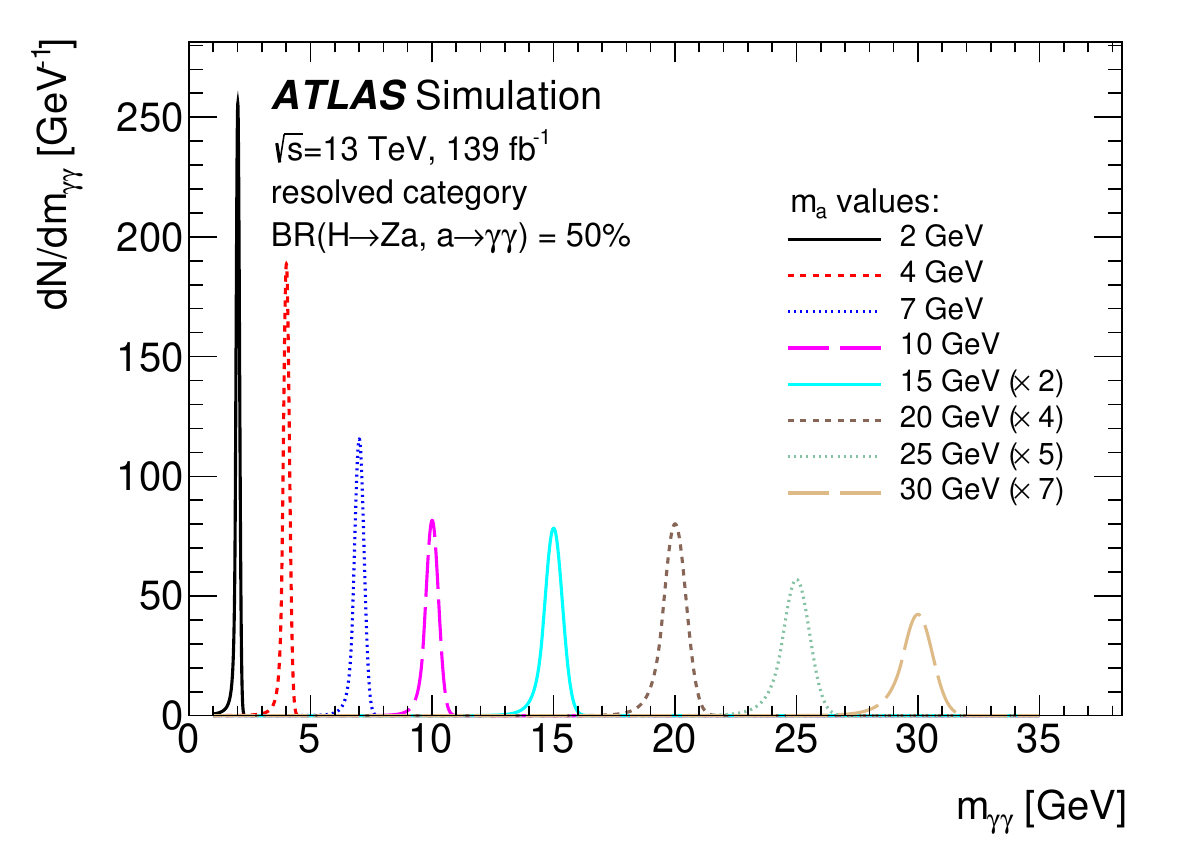}}\\
\subfloat[]{\label{fig:resolved:postfit}
\includegraphics[width=0.55\textwidth]{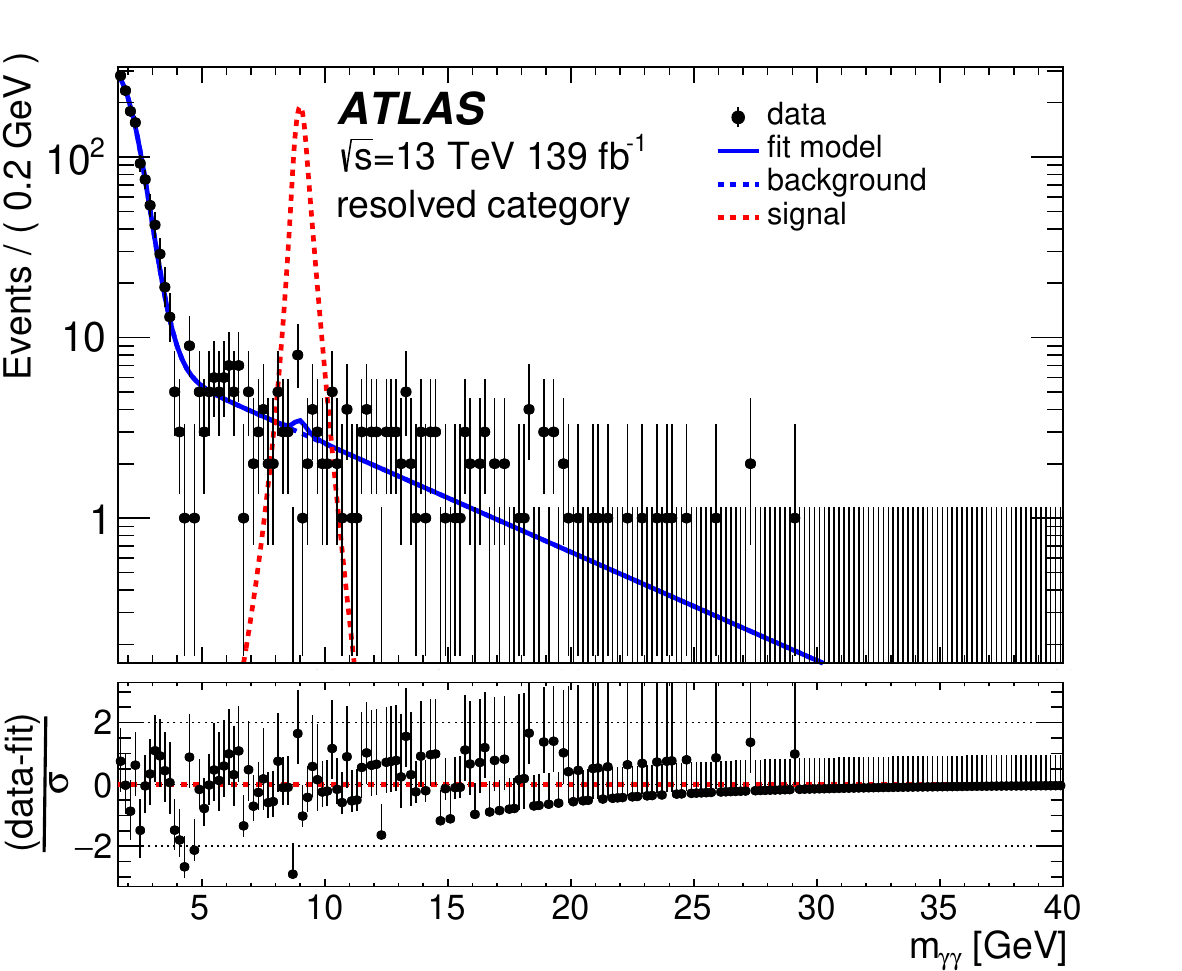}}
\caption{Distributions related to the resolved category:
(a) $m_{Z\gaga}$ distribution along with the range, indicated by vertical dashed lines, in which the resolved category events are found;
(b) signal \mgg distributions for a selection of $m_a$ values;
(c) \mgg post-fit distribution. Overlayed on (c)
for comparison is a signal distribution for $m_a = 9$~\GeV, and the signal plus background fit (solid blue) with its background component (dashed blue).
For (b) and (c) the branching ratio of the Higgs boson decay to $Za$ times the branching ratio $a\to\gamma\gamma$ is assumed to be 50\%.}
\label{fig:resolved}
\end{figure}

The events in the resolved category are required to have the invariant  mass of the $Z\gaga$ system,
$m_{Z\gaga}$, in the range 110--140~\GeV, so that they are
compatible with the 125~\GeV\ Higgs boson decay.
The $m_{Z\gaga}$ distributions for data and simulation are shown in Fig.~\ref{fig:resolved:mllgg}.
Here simulation is used to provide insight into the background composition
but is not employed for the final background estimation.
The main source of background is
$Z$+jets production, amounting to about 90\% of all background events.
Photons in this background mostly arise
from $\pi^0$ and other light meson decays. The remaining 10\% of
the background events are due to photons from \Zga production.
 
The final result of the search in the resolved category is extracted by
fitting the invariant mass of the $\gamma\gamma$ pair, \mgg.
For the signal, the \mgg distributions are modelled
with a parametric
function of which the parameters are interpolated to include signal
masses for which simulation is not available. A double-sided crystal
ball function~\cite{Oreglia:1980cs} is found to be an excellent
choice for the $m_{\gamma\gamma}$
distribution shape. Its parameters are interpolated to cover
any $a$ particle mass using linear functions. Examples
of signal \mgg distributions in the mass range used for
this category are shown in Fig.~\ref{fig:resolved:signal},
assuming a branching ratio of the Higgs boson decay to $Za$ times the branching ratio $a\to\gamma\gamma$ of 50\%.

The background distribution corresponds to
a smooth and falling spectrum for $m_a \geq 2.0$~\GeV.
This allows the background estimation to be performed with
a data-driven method in which continuum shape is parametrised
by an analytic function, which is derived in a control region
obtained by inverting the $m_{Z\gamma\gamma}$ requirement.
This method defines the region of interest
for the resolved category to start from $m_a = 2.0$~\GeV.
The parametric form of the background is:
$f^{bkg}(\mgg;a,b,c,d) = \frac{a}{1+e^{\frac{m_{\gamma\gamma}-b}{c}}} + (1-a)e^{-\frac{m_{\gamma\gamma}-20}{d}}$,
with four parameters $a$, $b$, $c$ and $d$.
It has two components: a Fermi-Dirac probability function to describe the bulk of the distribution,
and an exponential function to capture the tail end of the distribution.
This functional form is chosen based on its ability to adequately model the background
shape in several control regions across the full invariant mass range of the search.

The bias arising from the choice of the background  model is evaluated
by fitting the signal and background contributions to data that are known to have
no signal~\cite{ATL-PHYS-PUB-2020-028}.
In this way, the biases inherent to this method, which is known as
``spurious signal'', are estimated.
Templates are created using data or simulation in the control region,
and are used as probability distribution functions to generate
distributions with the expected
number of events in the signal region.
These distributions are fitted with a signal and background model
and the ``spurious signal'' is extracted. The median values of the
extracted signal for each $m_a$ assumption is used across all templates
to define an envelope that approximates the bias induced by the method.
This bias is included as a systematic uncertainty in this category.
Its value depends on the mass range, and the impact is limited as the size
is well below the statistical uncertainties in the signal extraction fit
for the entire mass range.

Post-fit $m_{\gamma\gamma}$ distributions with a binning of 200 MeV in the
signal region and including all the systematic uncertainties on the
background, as described earlier, and for signal (see Section~\ref{subsec:syst})
are shown in Fig.~\ref{fig:resolved:postfit}. In this figure,
a representative axion signal with a mass of $m_a=9$~\GeV\
and a branching ratio of the Higgs boson decay to $Za$ times the branching ratio $a\to\gamma\gamma$ of 50\% is shown.
 
\subsection{Merged category}\label{subsec:merged}

\begin{figure}
\centering
\subfloat[]{\label{fig:merged:a}
\includegraphics[width=0.45\textwidth]{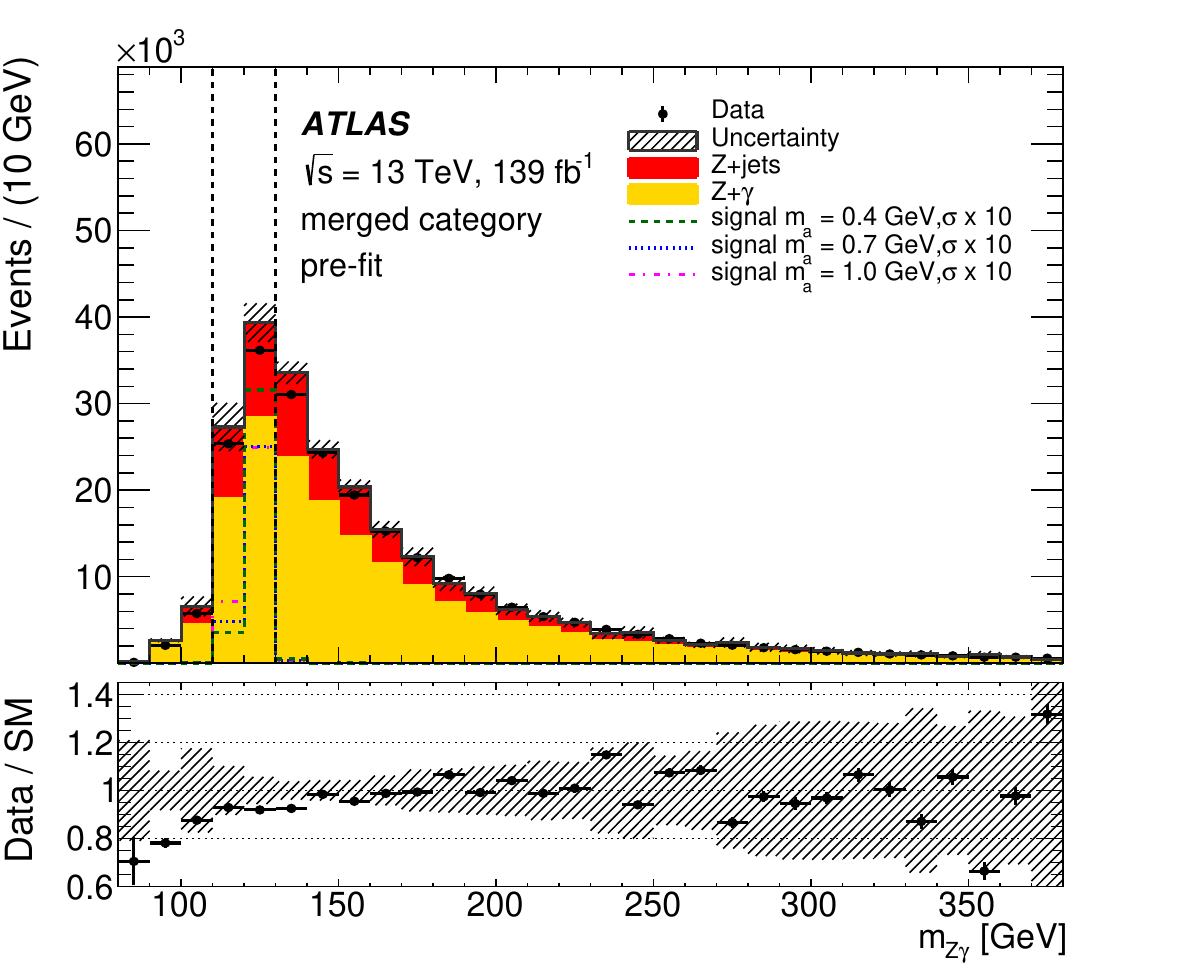}}
\subfloat[]{\label{fig:merged:b}
\includegraphics[width=0.45\textwidth]{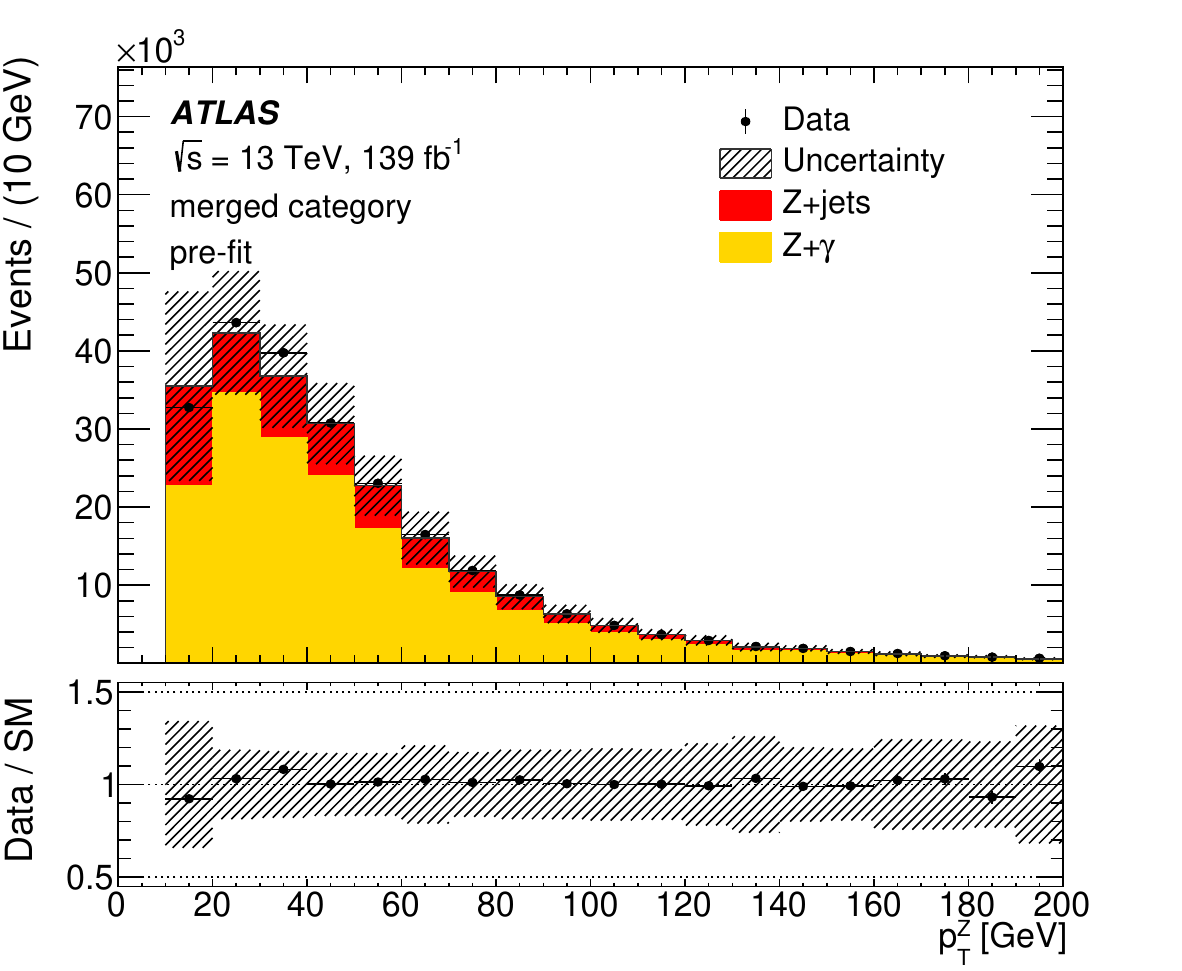}}  \\
\subfloat[]{\label{fig:merged:c}
\includegraphics[width=0.45\textwidth]{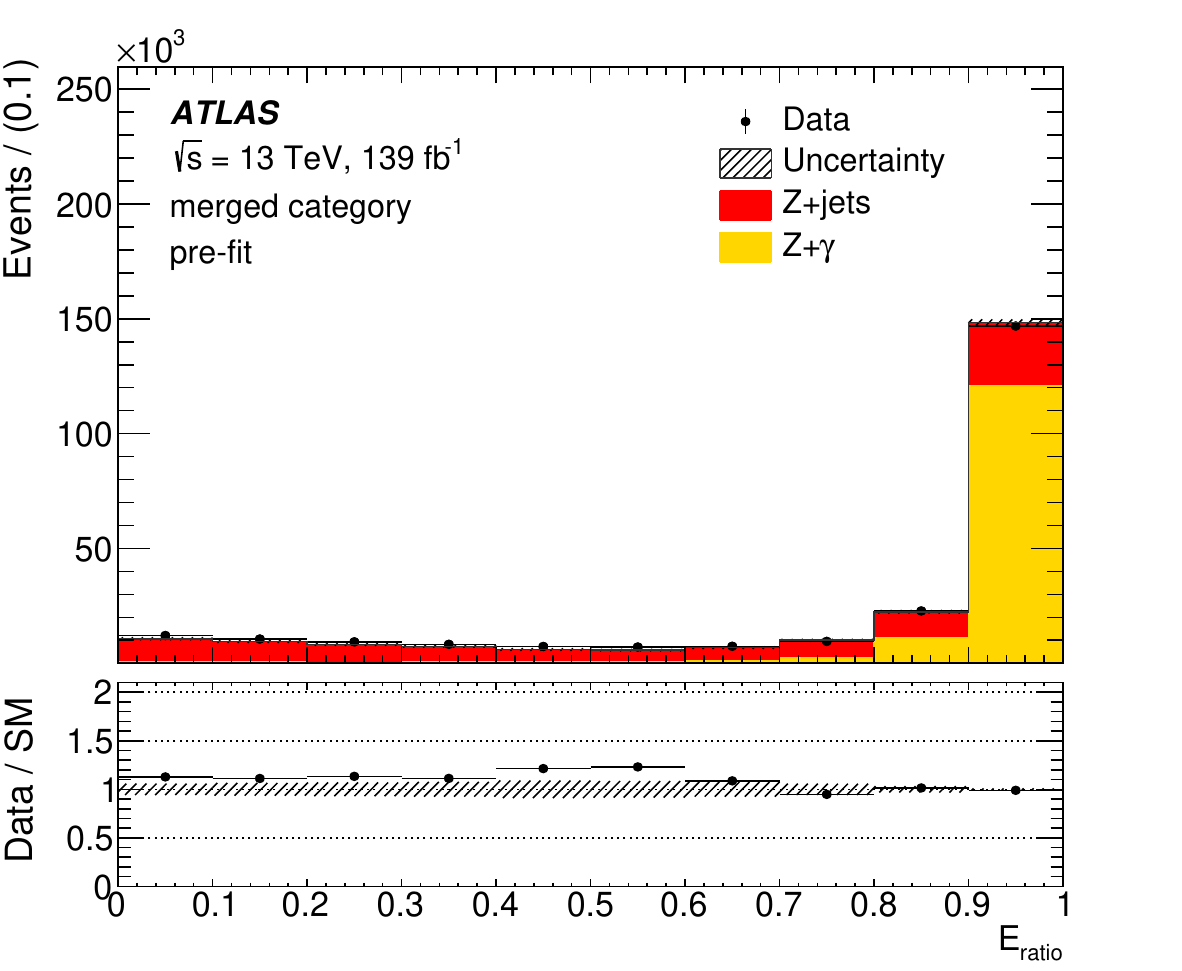}}
\subfloat[]{\label{fig:merged:d}
\includegraphics[width=0.45\textwidth]{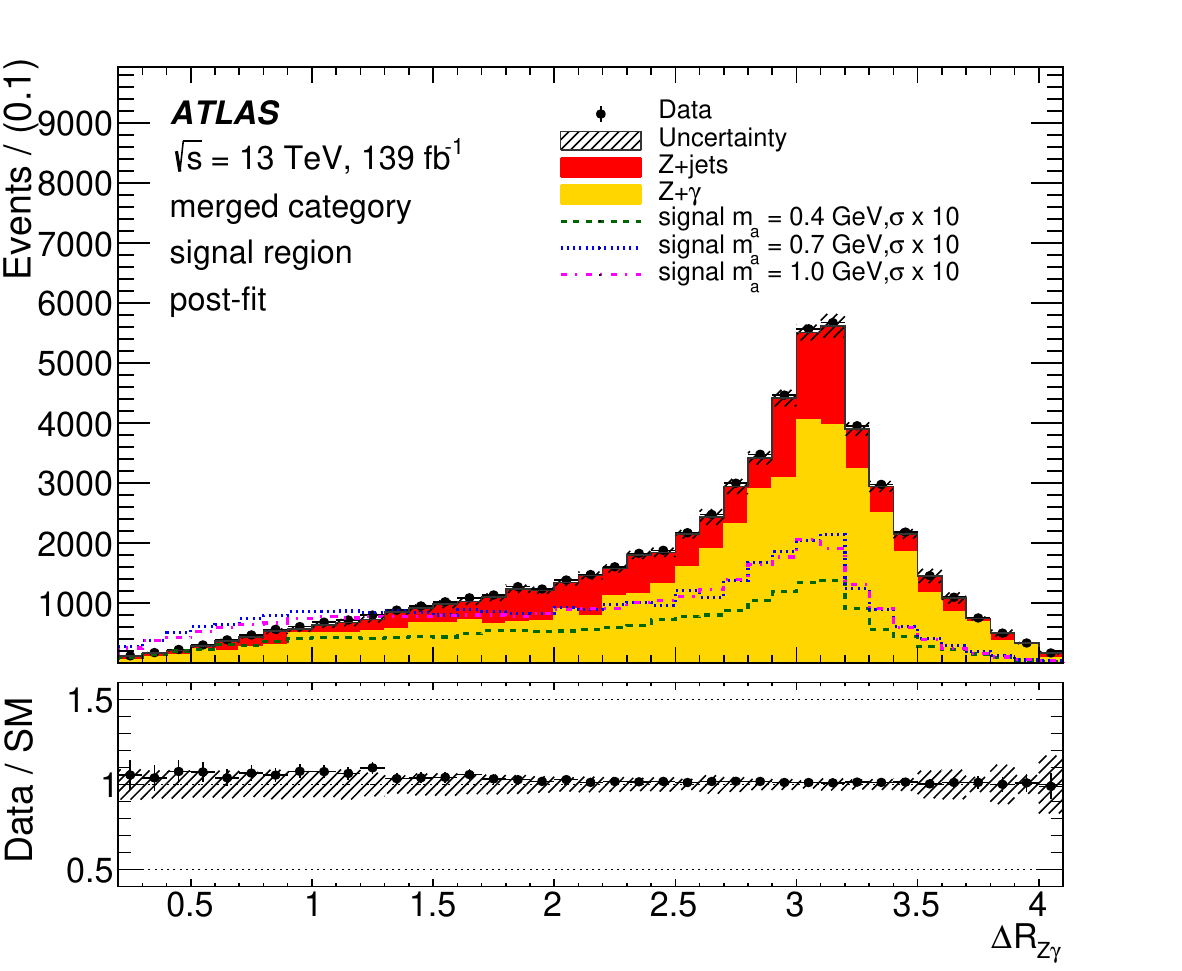}}
\caption{Distributions related to the merged category:
(a) $m_{Z\gamma}$ distribution along with the range, indicated by vertical dashed lines, in which the merged category events after the $E_{\text{ratio}}$ cut are found;
(b) \ptz distribution after categorisation, the $E_{\text{ratio}}$ cut and the corrections derived from the sideband control region;
(c) pre-fit distribution of the $E_{\text{ratio}}$ variable in the sideband  control region;
(d) post-fit final discriminating variable $\Delta R_{Z\gamma}$ in the signal region.
Signal distributions for $m_a$ values used in this category are overlayed for comparison, assuming
a branching ratio of the Higgs boson decay to $Za$ times the branching ratio $a\to\gamma\gamma$ of 100\%.
All the systematic uncertainties discussed in Section~\ref{subsec:syst} are included in these plots.
}
\label{fig:merged}
\end{figure}

The events in the merged category are required to have the mass of
the $Z\gamma$ system, $m_{Z\gamma}$ in the range 110--130~\GeV.
In addition, the photon must satisfy $E_{\text{ratio}} > 0.8$.
The quantity  $E_{\text{ratio}}$ is the ratio of the energy difference
between the maximum energy deposit
and the energy deposit in the secondary maximum in the photon cluster
to the sum of these energies~\cite{PERF-2017-02}. This variable
is very efficient in discriminating  photons against
jets mis-identified as photons.
The $E_{\text{ratio}} $ requirement reduces the background
from $Z+$jets by 60\%, leaving $Z+\gamma$ as the dominant background,
corresponding to about 75\% of the expected background events
in the signal region.
The $m_{Z\gamma}$ pre-fit distribution after the $E_{\text{ratio}}$ requirement
is shown in Fig.~\ref{fig:merged:a}.
 
The background estimation in the merged category uses simulation
with shape corrections derived from a control region that is composed of
events that fail the $m_{Z\gamma}$ requirement.
This control region is referred to as the ``sideband'' region.
The shape corrections are applied to $Z+\gamma$ and $Z+$jets events to correct for
differences between simulation and data for a number of chosen variables.
The sideband region primarily consists of an enriched population of $Z+\gamma$ events.
Conversely, a sample enriched in $Z+$jets events is obtained
by reversing the $E_{\text{ratio}}$ selection within the sideband.
The distribution of $E_{\text{ratio}} $ in this control region before the fit
is shown in Fig.~\ref{fig:merged:c}.

The $Z$+jets and $Z+\gamma$ modelling
is tested and corrected using data from the sideband control region.
In particular, corrections on the shape of the two-lepton system \pt,
$\pt^{Z}$, the two-lepton-plus-photon system \pt, $\pt^{Z\gamma}$,
the angular distance between the photon and the closest lepton,
$\Delta R_{\ell\gamma}$, and the photon \Et are individually derived
by comparing the shapes of these distributions in the sideband region
in data and in simulation.
The most relevant correction is for the shape of $\ptz$  and is shown pre-fit
after the $E_{\text{ratio}}$ requirement in Fig.~\ref{fig:merged:b}.

After applying all the corrections, the signal is extracted through a simultaneous fit of
two key observables: the angular distance between the two-lepton system and the $a \rightarrow \gamma\gamma$
candidate photon, $\Delta R_{Z\gamma}$, in the signal region; and the $E_\text{ratio}$ distribution in the sideband region.
This approach effectively constrains the normalisation of the $Z+$jets and \Zga backgrounds using the available data in the sideband region.
To ensure enhanced purity of the regions and minimise correlation between normalisation factors, the $E_\text{ratio}$ distribution is divided into two bins.
The first bin represents $E_\text{ratio} < 0.8$, while the second bin corresponds to $E_\text{ratio} > 0.8$.
In the signal region, as shown in Fig.~\ref{fig:merged:d} post-fit, the background exhibits a pronounced peak for large $\Delta R_{Z\gamma}$ values,
while the signal shape demonstrates a long tail for lower $\Delta R_{Z\gamma}$ values.

\subsection{Systematic uncertainties} \label{subsec:syst}
Systematic effects from both experimental and theoretical sources affect
the signal and background estimation
used in deriving the results of the analysis.
These may change the number of expected events and the shape of the fitted distributions. The following uncertainties are included in the signal distributions
of both the merged and resolved categories and in the background estimation
of the merged category. Dedicated systematic uncertainties coming from the background
estimation in the resolved category and the $Z$+jets or \Zga modelling in the
merged category are described in Sections~\ref{subsec:resolved}
and~\ref{subsec:merged}, respectively.
 
Background shape uncertainties of the MC distributions in the merged category are evaluated from the shape corrections applied to the samples.
The full size of the shape correction is taken as an estimate of the associated systematic uncertainty.
The impact of these uncertainties on the final result applied to the $Z+\gamma$ process in the $\Delta R_{Z\gamma}$ distribution,
is found to be less than 0.5\%.
The most significant impact is from the shape of $\ptz$ in the $Z$+jets process. However the impact on the final result is small as the $Z$+jets is a
sub-dominant background.
 
Normalisation uncertainties related to modelling corrections of
electrons and muons in the simulation with respect to data
as well as the trigger efficiencies are considered. The largest impacts are
from the electron identification and the muon reconstruction,
which are found, however,  to be less than 1.5\% and 0.5\% respectively across all
simulated samples.
 
Uncertainties on the data to simulation correction factors for photon identification
and isolation efficiencies are evaluated as the difference between the correction
factors applied or not, and they are found to be at maximum 0.3\% and 6.5\%
respectively for the signal, and 0.3\% and 2.0\% for the background processes.
The photon energy scale and resolution uncertainties are found to be
negligible and thus not included in the analysis.
 
A pile-up modelling uncertainty is assigned to account for the
difference between the predicted and measured inelastic $pp$ cross-sections~\cite{STDM-2015-05}.
This uncertainty is evaluated to have a less than 1.5\% effect on the event
yield across all signal samples and less than 15\% for the background processes.
 
Modelling uncertainty arises from various sources, including limited MC sample statistics,
uncertainty in the renormalisation and factorisation scales, the choice of MC generator
for both the signal and backgrounds, and the uncertainty in the parton distribution functions.
These are evaluated to be less than 10\% and 2\% for the gluon--gluon fusion and the VBF processes, respectively.
 
The uncertainty in the integrated luminosity of the combined dataset from 2015 to 2018 is 1.7\%,
using the same methodology as in Ref.~\cite{DAPR-2021-01}, and obtained using the LUCID-2
detector~\cite{LUCID2} for the primary luminosity measurements, complemented by measurements
using the inner detector and calorimeters.
 
The background parametrisation in the resolved category is data-driven
and extracted from the fit in the sideband region,
thus only systematic uncertainties related to this fit and the signal modeling are considered.
The systematic variation of the fit parameters is found to have a
negligible impact on the fit result.
The bias from the choice of the background model is evaluated with
the ``spurious signal'' method, as discussed in Section~\ref{subsec:resolved}.
 
All systematic uncertainties are implemented through nuisance parameters
in the likelihood function used to fit the data and are profiled in the
final fit, as described in more detail in Section~\ref{sec:results}.
A summary of the impact on the limit for the merged category of each systematic uncertainty is
reported in Table~\ref{tab:sys} for a signal-plus-background fit
assuming a signal model with an axion mass of 1~\gev. These are computed by repeating the
fit procedure after fixing the nuisance parameter under evaluation to its best-fit value, with
all other parameters floating. The impact is then obtained by comparing the quadratic difference between
the regular fit and the alternative fit with the parameter fixed.
 
\begin{table}
\centering
\small
\begin{tabular}{lc}
\noalign{\smallskip}\hline\noalign{\smallskip}
Source                                       &  $1\sigma$ Uncertainty\\
\noalign{\smallskip}\hline\noalign{\smallskip}
Total                                        & 28\%    \\
Data statistical unc.                        & 1.4\%       \\
MC statistical unc.                          & 26\%       \\
Total background                             & 5.6\%       \\
\noalign{\smallskip}\hline\noalign{\smallskip}
Signal modelling                             & 5.2\%   \\
Pile-up                                      & 0.2\%   \\
Electron identification                      & 0.4\%      \\
Photon isolation                             & 0.9\%     \\
Muon trigger                                 & 0.6\%      \\
Muon reconstruction                          & 0.3\%      \\
Muon isolation                               & 0.3\%  \\
$Z+\gamma$ normalisation                     & 0.8\%   \\
$Z$ + jets normalisation                       & 0.9\%    \\
$Z+\gamma$ corrections                       & 0.3\%     \\
$Z$ + jets corrections                         & 1.0\%  \\
\noalign{\smallskip}\hline\noalign{\smallskip}
\end{tabular}
\caption{
The contributions to the 68\% confidence interval of the fitted signal strength from different sources of systematic uncertainties, for a  signal-plus-background fit for the merged category assuming a signal model with an axion mass of 1~\gev.
The evaluation is performed by fixing the parameter related to the tested source of systematic uncertainty to the best-fit value and redoing the fit with all other parameters floating. The impact is then obtained by comparing the quadratic difference between the default fit and the new model with the parameter fixed.
The individual uncertainties can be correlated, so do not necessarily add up quadratically to the total uncertainty. The percentages show the size of the uncertainty relative to the parameter of interest.}
\label{tab:sys}
\end{table}
\clearpage


\section{Results}
\label{sec:results}

The statistical analysis of the data uses binned maximum-likelihood
fits of the $m_{\gamma\gamma}$ and $\Delta R_{Z\gamma}$
distributions in the resolved and merged categories, respectively.
The search in the observed $m_{\gamma\gamma}$ distribution is performed in
the 2 to 33 GeV mass range with a 200 MeV step, smaller than the $a$ particle invariant mass resolution.
The systematic uncertainties discussed in Section~\ref{subsec:syst} are
accounted for in the fit by means of nuisance parameters
constrained by Gaussian penalty terms in the likelihood function.
In the absense of signal, expected and observed 95\%
CL exclusion limits on the
branching ratio of the Higgs boson decay to $Za$ times the branching ratio $a\to\gamma\gamma$
are computed as a function of $m_a$ using the CL$_s$~\cite{Read:2002hq} modified
frequentist approach.
The exclusion limits are computed using the asymptotic approximation~\cite{Cowan:2010js}.
 
\begin{figure}[b]
\centering
\subfloat[]{ \label{fig:results:1dLimit}
\includegraphics[height=6cm]{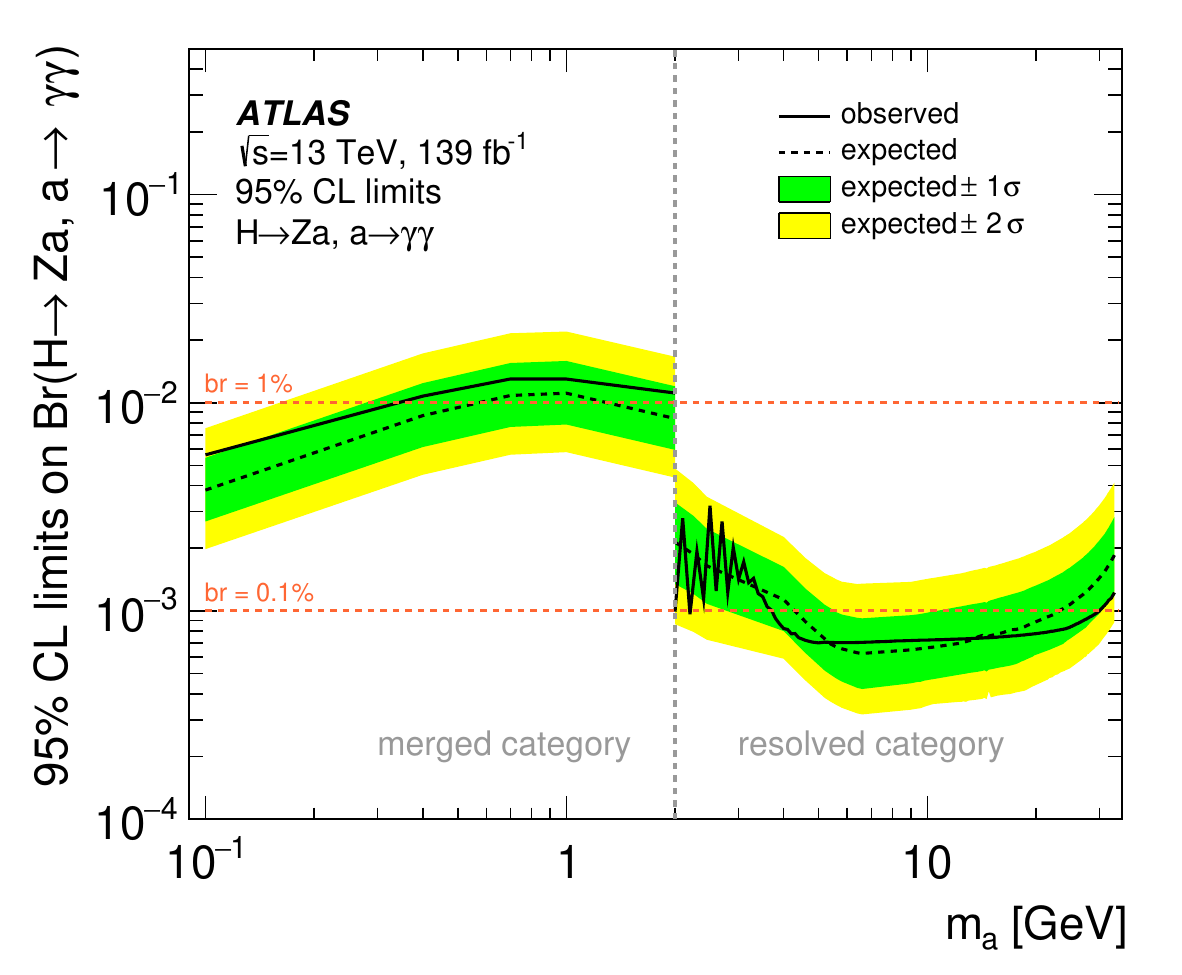}}
\subfloat[]{ \label{fig:results:2dLimit}
\includegraphics[height=6.2cm]{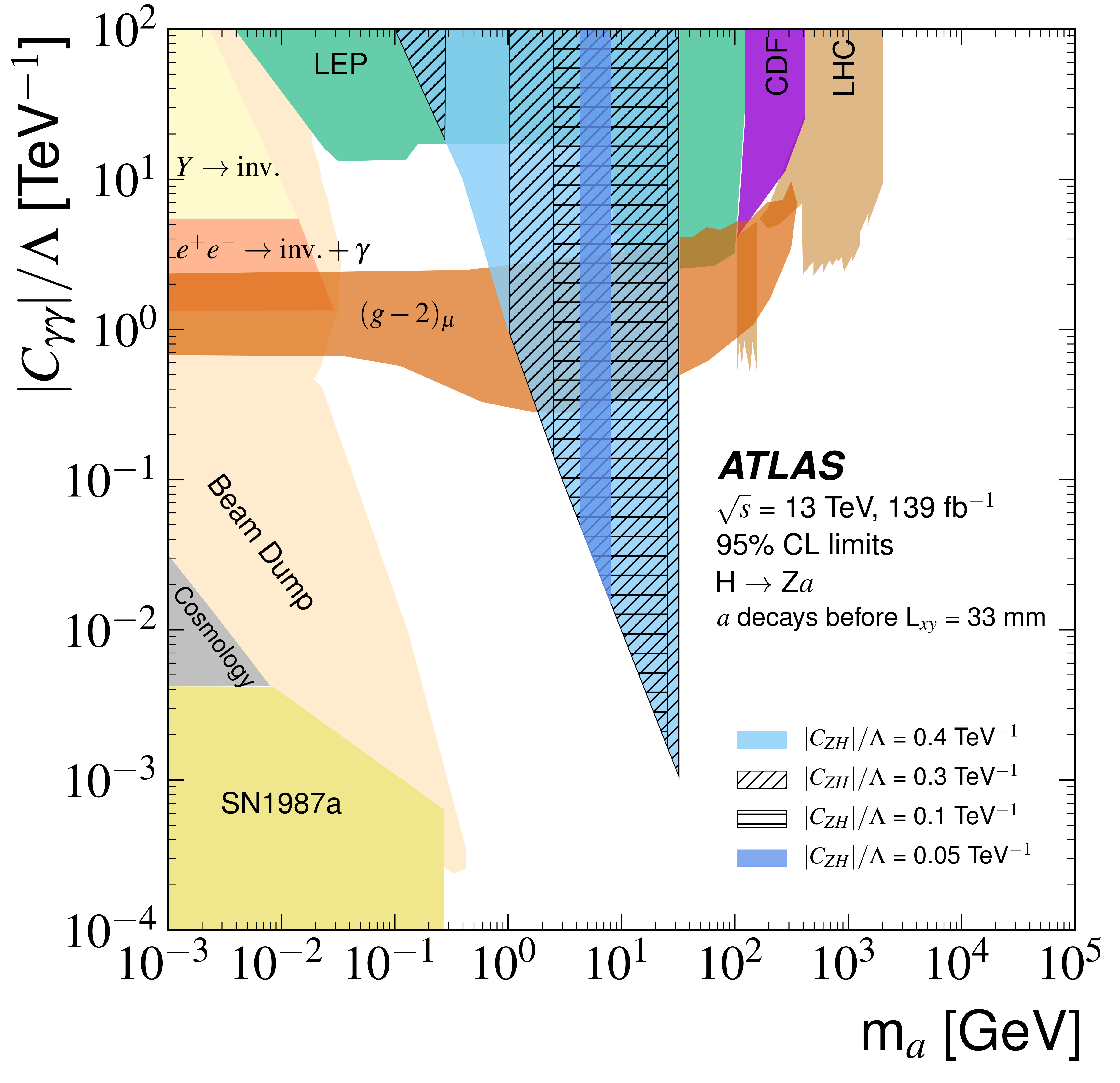}}
\caption{(a) Expected and observed 95\% CL upper limits on
the branching ratio of the Higgs boson decay to $Za$ times the branching ratio $a\to\gamma\gamma$
as a function of the $a$ particle mass in the merged ($m_a \le 2~ \GeV$) and the resolved ($m_a > 2~ \GeV$) categories.
(b) ATLAS observed 95\% CL exclusion contours limits in terms of the ALP mass and its effective coupling to photons, $|C_{\gamma\gamma}|/\Lambda$, for different values of the Higgs coupling to $Za$, $|C_{ZH}|/\Lambda$.
The overlaid contour limits from other direct experimental searches are shown as well. The collider bounds (LHC, LEP, CDF) are displayed at 95\% CL, while the remaining bounds (SN1987a, Cosmology and Beam Dump) are presented at 90\% CL. The red band shows the preferred parameter space where the $(g-2)\mu$ anomaly can be explained at 95\% CL. These contours are adapted from Refs.~\cite{Jaeckel:2015jla,Bauer:2017nlg}.}
\label{fig:limit}
\end{figure}

The 95\% CL upper limits for both categories are shown in Fig.~\ref{fig:results:1dLimit}.
The validity of the resolved category is limited to masses larger than 2 GeV and is complementary to the merged category.
The search excludes a range of branching ratios of
the branching ratio of the Higgs boson decay to $Za$ times the branching ratio $a\to\gamma\gamma$ from 0.08\% to 2\%,
depending on the assumed $m_a$. No significant excesses are observed.
 
The upper limits are also interpreted in the context of ALP models~\cite{Bauer:2017ris},
in terms of the ALP mass and its effective coupling to photons, $|C_{\gaga}|/\Lambda$. These are presented in
Fig.~\ref{fig:results:2dLimit} for different values of the Higgs coupling to $Za$, $|C_{ZH}|/\Lambda$),
ranging from 0.05~\TeV$^{-1}$~to~0.4~\TeV$^{-1}$.
Depending on the  $|C_{\gaga}|/\Lambda$ coupling values the ALP may have a sizeable lifetime resulting
in displaced decays. The results of the analysis are fully accurate only for prompt axions decays,
and only models with ALPs decaying up to the first ATLAS pixel layer ($L_{xy}=33\ \mathrm{mm}$) are considered. In the same figure, 90\% CL exclusion regions on the ALP parameter
space from direct experimental searches are
also shown for comparison, with these contours being
adapted from Refs.~\cite{Jaeckel:2015jla,Bauer:2017nlg}.
 
\clearpage
\section{Conclusion}
\label{sec:conclusion}
 
Data recorded by the ATLAS experiment at the LHC corresponding to
an integrated luminosity of 139~fb$^{-1}$ from proton--proton collisions
at a centre-of-mass energy of 13~\TeV, are used to search for a rare
decay of the Higgs boson to a $Z$ boson and an axion
particle $a$, with a mass between 0.1~\GeV~and
33~\GeV.
The $Z$ boson is reconstructed using an electron or muon pair, while
the particle $a$ candidate is reconstructed from a pair of photons.
The analysis accounts for the cases in which both photons are
close enough to be reconstructed as a single photon in the detector,
or topologies where they are reconstructed as two
separate photons. No significant deviations with respect to
the SM predictions are observed and
upper limits are set on the branching ratio
of the Higgs boson decay to $Za$ times the branching ratio $a\to\gamma\gamma$, ranging from from 0.08\% to 2\%
depending on the mass of the $a$ particle.
This is one of the first direct limits on the Higgs boson branching fraction
into a $Z$ boson and an ALP decaying to a pair of photons,
reaching sensitivity to parameter space that was not accessible
by previous searches.


\section*{Acknowledgements}


We thank CERN for the very successful operation of the LHC, as well as the
support staff from our institutions without whom ATLAS could not be
operated efficiently.
 
We acknowledge the support of
ANPCyT, Argentina;
YerPhI, Armenia;
ARC, Australia;
BMWFW and FWF, Austria;
ANAS, Azerbaijan;
CNPq and FAPESP, Brazil;
NSERC, NRC and CFI, Canada;
CERN;
ANID, Chile;
CAS, MOST and NSFC, China;
Minciencias, Colombia;
MEYS CR, Czech Republic;
DNRF and DNSRC, Denmark;
IN2P3-CNRS and CEA-DRF/IRFU, France;
SRNSFG, Georgia;
BMBF, HGF and MPG, Germany;
GSRI, Greece;
RGC and Hong Kong SAR, China;
ISF and Benoziyo Center, Israel;
INFN, Italy;
MEXT and JSPS, Japan;
CNRST, Morocco;
NWO, Netherlands;
RCN, Norway;
MEiN, Poland;
FCT, Portugal;
MNE/IFA, Romania;
MESTD, Serbia;
MSSR, Slovakia;
ARRS and MIZ\v{S}, Slovenia;
DSI/NRF, South Africa;
MICINN, Spain;
SRC and Wallenberg Foundation, Sweden;
SERI, SNSF and Cantons of Bern and Geneva, Switzerland;
MOST, Taiwan;
TENMAK, T\"urkiye;
STFC, United Kingdom;
DOE and NSF, United States of America.
In addition, individual groups and members have received support from
BCKDF, CANARIE, Compute Canada and CRC, Canada;
PRIMUS 21/SCI/017 and UNCE SCI/013, Czech Republic;
COST, ERC, ERDF, Horizon 2020, ICSC-NextGenerationEU and Marie Sk{\l}odowska-Curie Actions, European Union;
Investissements d'Avenir Labex, Investissements d'Avenir Idex and ANR, France;
DFG and AvH Foundation, Germany;
Herakleitos, Thales and Aristeia programmes co-financed by EU-ESF and the Greek NSRF, Greece;
BSF-NSF and MINERVA, Israel;
Norwegian Financial Mechanism 2014-2021, Norway;
NCN and NAWA, Poland;
La Caixa Banking Foundation, CERCA Programme Generalitat de Catalunya and PROMETEO and GenT Programmes Generalitat Valenciana, Spain;
G\"{o}ran Gustafssons Stiftelse, Sweden;
The Royal Society and Leverhulme Trust, United Kingdom.
 
The crucial computing support from all WLCG partners is acknowledged gratefully, in particular from CERN, the ATLAS Tier-1 facilities at TRIUMF (Canada), NDGF (Denmark, Norway, Sweden), CC-IN2P3 (France), KIT/GridKA (Germany), INFN-CNAF (Italy), NL-T1 (Netherlands), PIC (Spain), ASGC (Taiwan), RAL (UK) and BNL (USA), the Tier-2 facilities worldwide and large non-WLCG resource providers. Major contributors of computing resources are listed in Ref.~\cite{ATL-SOFT-PUB-2023-001}.


\printbibliography
 
\clearpage
 
\begin{flushleft}
\hypersetup{urlcolor=black}
{\Large The ATLAS Collaboration}

\bigskip

\AtlasOrcid[0000-0002-6665-4934]{G.~Aad}$^\textrm{\scriptsize 102}$,
\AtlasOrcid[0000-0002-5888-2734]{B.~Abbott}$^\textrm{\scriptsize 120}$,
\AtlasOrcid[0000-0002-1002-1652]{K.~Abeling}$^\textrm{\scriptsize 55}$,
\AtlasOrcid[0000-0001-5763-2760]{N.J.~Abicht}$^\textrm{\scriptsize 49}$,
\AtlasOrcid[0000-0002-8496-9294]{S.H.~Abidi}$^\textrm{\scriptsize 29}$,
\AtlasOrcid[0000-0002-9987-2292]{A.~Aboulhorma}$^\textrm{\scriptsize 35e}$,
\AtlasOrcid[0000-0001-5329-6640]{H.~Abramowicz}$^\textrm{\scriptsize 151}$,
\AtlasOrcid[0000-0002-1599-2896]{H.~Abreu}$^\textrm{\scriptsize 150}$,
\AtlasOrcid[0000-0003-0403-3697]{Y.~Abulaiti}$^\textrm{\scriptsize 117}$,
\AtlasOrcid[0000-0002-8588-9157]{B.S.~Acharya}$^\textrm{\scriptsize 69a,69b,m}$
% ...
\bigskip
\\

% ...
$^{29}$Physics Department, Brookhaven National Laboratory, Upton NY; United States of America.\\
% ...
$^{35}$$^{(a)}$Facult\'e des Sciences Ain Chock, R\'eseau Universitaire de Physique des Hautes Energies - Universit\'e Hassan II, Casablanca;$^{(b)}$Facult\'{e} des Sciences, Universit\'{e} Ibn-Tofail, K\'{e}nitra;$^{(c)}$Facult\'e des Sciences Semlalia, Universit\'e Cadi Ayyad, LPHEA-Marrakech;$^{(d)}$LPMR, Facult\'e des Sciences, Universit\'e Mohamed Premier, Oujda;$^{(e)}$Facult\'e des sciences, Universit\'e Mohammed V, Rabat;$^{(f)}$Institute of Applied Physics, Mohammed VI Polytechnic University, Ben Guerir; Morocco.\\
% ...
$^{49}$Fakult\"{a}t Physik , Technische Universit{\"a}t Dortmund, Dortmund; Germany.\\
% ...
$^{55}$II. Physikalisches Institut, Georg-August-Universit\"{a}t G\"ottingen, G\"ottingen; Germany.\\
% ...
$^{69}$$^{(a)}$INFN Gruppo Collegato di Udine, Sezione di Trieste, Udine;$^{(b)}$ICTP, Trieste;$^{(c)}$Dipartimento Politecnico di Ingegneria e Architettura, Universit\`a di Udine, Udine; Italy.\\
% ...
$^{102}$CPPM, Aix-Marseille Universit\'e, CNRS/IN2P3, Marseille; France.\\
% ...
$^{117}$Department of Physics, New York University, New York NY; United States of America.\\
% ...
$^{120}$Homer L. Dodge Department of Physics and Astronomy, University of Oklahoma, Norman OK; United States of America.\\
% ...
$^{150}$Department of Physics, Technion, Israel Institute of Technology, Haifa; Israel.\\
$^{151}$Raymond and Beverly Sackler School of Physics and Astronomy, Tel Aviv University, Tel Aviv; Israel.\\
% ...
$^{a}$ Also Affiliated with an institute covered by a cooperation agreement with CERN.\\
$^{b}$ Also at An-Najah National University, Nablus; Palestine.\\
$^{c}$ Also at Borough of Manhattan Community College, City University of New York, New York NY; United States of America.\\
$^{d}$ Also at Center for High Energy Physics, Peking University; China.\\
$^{e}$ Also at Center for Interdisciplinary Research and Innovation (CIRI-AUTH), Thessaloniki; Greece.\\
$^{f}$ Also at Centro Studi e Ricerche Enrico Fermi; Italy.\\
% ...
$^{m}$ Also at Department of Physics, King's College London, London; United Kingdom.\\
% ...

\end{flushleft}


\end{document}